\documentclass[12pt]{article}
\usepackage{graphicx}
\usepackage{epsfig}
\usepackage{epstopdf}
\DeclareGraphicsExtensions{.pdf,.eps,.png,.jpg,.mps}

\def\lsim{\mathrel{\rlap{\lower3pt\hbox{\hskip0pt$\sim$}}
     \raise1pt\hbox{$<$}}}         
\def\gsim{\mathrel{\rlap{\lower4pt\hbox{\hskip1pt$\sim$}}
     \raise1pt\hbox{$>$}}}         

\usepackage{amsmath}
\usepackage{amsfonts}

\begin{document}
\begin{titlepage}

\centerline{\Large \bf Factor Models for Alpha Streams}
\medskip

\centerline{Zura Kakushadze$^\S$$^\dag$$^\ddag$\footnote{\, Zura Kakushadze, Ph.D., is the President of Quantigic$^\circledR$ Solutions LLC, an Adjunct Professor at the University of Connecticut, and a Full Professor at Free University of Tbilisi. Email: \tt zura@quantigic.com}}
\bigskip

\centerline{\em $^\S$ Quantigic$^\circledR$ Solutions LLC}
\centerline{\em 1127 High Ridge Road \#135, Stamford, CT 06905\,\,\footnote{\, DISCLAIMER: This address is used by the corresponding author for no
purpose other than to indicate his professional affiliation as is customary in
publications. In particular, the contents of this paper
are not intended as an investment, legal, tax or any other such advice,
and in no way represent views of Quantigic® Solutions LLC,
the website \underline{www.quantigic.com} or any of their other affiliates.
}}
\centerline{\em $^\dag$ Department of Physics, University of Connecticut}
\centerline{\em 1 University Place, Stamford, CT 06901}
\centerline{\em $^\ddag$ Free University of Tbilisi, Business School \& School of Physics}
\centerline{\em 240, David Agmashenebeli Alley, Tbilisi, 0159, Georgia}
\medskip
\centerline{(June 12, 2014; revised: September 3, 2014)}

\bigskip
\medskip

\begin{abstract}
{}We propose a framework for constructing factor models for alpha streams. Our motivation is threefold. 1) When the number of alphas is large, the sample covariance matrix is singular. 2) Its out-of-sample stability is challenging. 3) Optimization of investment allocation into alpha streams can be tractable for a factor model alpha covariance matrix. We discuss various risk factors for alphas such as: style risk factors; cluster risk factors based on alpha taxonomy; principal components; and also using the underlying tradables (stocks) as alpha risk factors, for which computing the factor loadings and factor covariance matrices does not involve any correlations with alphas, and their number is much larger than that of the relevant principal components. We draw insight from stock factor models, but also point out substantial differences.
\end{abstract}
\end{titlepage}

\newpage

\section{Motivation and Summary}

{}It appears to be a natural tendency that the number of investable alpha streams\footnote{\, For a partial list of hedge fund literature, see, {\em e.g.}, \cite{HF1}-\cite{HF20} and references therein.} grows with time. In the olden days, alphas, which can be thought of as sets of instructions for taking predefined positions in underlying tradables at specified times, were built ``by hand". Nowadays, many thousands of alpha streams can be datamined in an automated fashion. With that comes an ``embarrassment of the riches" of sorts -- there are too many alpha streams and comparatively too few historical observations. As a result, making predictions about future performance of these alpha streams becomes challenging, not only in terms of out-of-sample stability, but also computing the alpha covariance matrix based on the alpha stream time series -- the sample covariance matrix is badly singular, precisely due to too few observations.

{}As a result, allocating investment into a large number of alpha streams, {\em i.e.}, computing optimal weights for such allocation, becomes nontrivial. Even if one employs the simplest optimization criterion and maximizes the Sharpe ratio of the combined alpha stream portfolio, one runs into an issue: this optimization requires inverting the alpha covariance matrix, which is singular. And even if one somehow regularizes the covariance matrix, it is not all that stable out-of-sample. The question we ponder in this note is how to approach this issue in a systematic way.

{}We look to history for insight. When the number of underlying tradables -- stocks -- became too large, one had to deal with a conceptually similar problem. To reliably compute a sample covariance matrix for stock (daily) returns, one would need a prohibitively large number of observations (trading days),\footnote{\, There is always the issue of what to do with new tickers that have little to no history. This is not what we refer to here.} even for a universe of 2,000-2,500 tickers. And even if that much history existed, going back a decade or longer makes little sense for many practical applications, because the relevant time horizons are much shorter, and many strategies are much shorter lived. A way around this difficulty is to employ a multi-factor risk model,\footnote{\, For a partial list of factor model and related literature, see, {\em e.g.}, \cite{FM1}-\cite{FM40} and references therein.} where one assumes that stock returns have some intrinsic specific risk, which must be measured empirically, plus factor risk, which is a linear combination of the underlying risk factors,\footnote{\, Examples of such risk factors are momentum, size, liquidity, volatility, growth, value, {\em etc.} (style risk factors), exposures to (sub-)industries (industry risk factors), principal components (``beta"-like risk factors), {\em etc.}} whose number $F_S$ is much smaller than then number of stocks $N_S$. The correlations between stocks then are attributed solely to their exposure to these risk factors, so the off-diagonal elements of the stock covariance matrix are determined by the $F_S \times F_S$ factor covariance matrix, computing which requires many fewer observations than computing the $N_S\times N_S$ stock covariance matrix, and it is also expected to be much more stable out-of-sample for a suitably chosen set of risk factors.

{}One can do essentially the same for alphas -- build {\em factor models for alpha streams}. Out of thousands of alphas one may construct, many are closely correlated to each other for a variety of reasons, including how they are constructed. However, there are also substantial differences between alpha factor models and stock factor models. The purpose of this note is to set forth a framework for constructing factor models for alpha streams and discuss various approaches and intricacies arising therein.

{}One additional motivation for factor models for alpha streams is that, when a number of alpha streams is traded on the same execution platform, allocation weights\footnote{\, For a partial list of portfolio selection literature, see, {\em e.g.}, \cite{PO1}-\cite{PO40} and references therein.} are allowed to be negative. Furthermore, if trades are crossed between different alphas, portfolio turnover reduces \cite{OD, SpMod, FMT}, further complicating the weight optimization problem. With these additional challenges, the investment allocation problem into alpha streams becomes rather difficult to tackle for a general alpha covariance matrix,\footnote{\, Even if it is made positive-definite via a deformation (see, {\em e.g.}, \cite{RJ, SpMod}).} especially once linear and nonlinear costs are added and/or optimization criteria beyond maximizing the Sharpe ratio are considered,\footnote{\, {\em E.g.}, maximizing P\&L with volatility bounded from above.} but can be tractable if the alpha covariance matrix has a factor model form \cite{AlphaOpt, AlphaWeights}.

{}The remainder of this note is organized as follows. In Section \ref{sec.setup} we set up our notations. In Section \ref{sec.sing} we discuss optimization for singular alpha covariance matrix via its regularization and how it reduces to a (generalized) weighted regression in the singular limit. In Section \ref{fac.mod} we set forth the framework for factor models for alpha streams. We discuss style risk factors for alphas, cluster risk factors based on alpha taxonomy and associated issues, and principal component risk factors that arise in the regression limit of Section \ref{sec.sing}. We then discuss using underlying tradables as risk factors for alphas and how to compute the corresponding factor loadings matrices. Concluding remarks are in Section \ref{concl}. Appendix \ref{app.a} discusses some aspects of capacity, which is one of the style factors one may choose to use. Appendix \ref{app.b} discusses some aspects of constructing the factor covariance matrix and specific risk.

\section{Definitions and Setup}\label{sec.setup}

{}We have $N$ alphas $\alpha_i$, $i=1,\dots,N$. Each alpha is actually a time series $\alpha_i(t_s)$, $s=0,1,\dots,M$, where $t_0$ is the most recent time. Below $\alpha_i$ refers to $\alpha_i(t_0)$.

{}Let $C_{ij}$ be the covariance matrix of the $N$ time series $\alpha_i(t_s)$. Let $\Psi_{ij}$ be the corresponding correlation matrix, {\em i.e.},
\begin{equation}
 C_{ij} = \sigma_i~\sigma_j~\Psi_{ij}
\end{equation}
where $\Psi_{ii} = 1$.

{}If $M\ll N$, which is the case in most practical applications with $N\gg 1$, then $C_{ij}$ is (nearly) degenerate with $M$ ``large" eigenvalues and the remainder having ``small" values, which can be positive or negative. These small values are zeros distorted by computational rounding.\footnote{\, Actually, this assumes that there are no N/As in any of the alpha time series. If some or all alpha time series contain N/As in non-uniform manner and the correlation matrix is computed by omitting such pair-wise N/As, then the resulting correlation matrix may have negative eigenvalues that are not ``small" in the sense used above, {\em i.e.}, they are not zeros distorted by computational rounding. For the sake of simplicity, here we assume that there are no N/As.\label{corrNAs}}

\subsection{Alpha Weight Optimization}

{}Suppose we wish to allocate investment $I$ into our $N$ alphas. We need to find the alpha weights $w_i$ such that
\begin{equation}\label{w}
 \sum_{i=1}^N \left|w_i\right| = 1
\end{equation}
where the modulus accounts for the fact that some weighs can be negative if the alphas are traded on the same execution platform. For the sake of simplicity, let us assume no transaction costs -- they are not important for point we are trying to arrive at here. Portfolio P\&L, volatility and Sharpe ratio are given by
\begin{eqnarray}
 &&P = I~\sum_{i=1}^N \alpha_i~w_i\\
 &&R = I~\left({\sum_{i,j=1}^N C_{ij}~w_i~w_j}\right)^{1\over 2}\\
 &&S = {P \over R}
\end{eqnarray}
The simplest weight optimization criterion is to maximize the Sharpe ratio:
\begin{equation}
 S\rightarrow \mbox{max}
\end{equation}
If the covariance matrix $C$ is nonsingular, then the Sharpe ratio is maximized by the following alpha weights $w_i$:
\begin{equation}\label{opt}
 w_i = \xi~\sum_{j=1}^N C^{-1}_{ij}~\alpha_j
\end{equation}
where $C^{-1}_{ij}$ is the inverse of $C_{ij}$, and $\xi$ is a normalization constant fixed by (\ref{w}).

\section{Singular Covariance Matrix}\label{sec.sing}

{}When $C$ is singular, one can regularize it by deforming it: $C\rightarrow \Gamma$. Such regularization can be parameterized as follows:
\begin{equation}\label{Gamma}
 \Gamma \equiv C + \epsilon~\Delta
\end{equation}
where $\epsilon$ is a regularization parameter ($\Gamma\rightarrow C$ when $\epsilon\rightarrow 0$), and $\Delta_{ij}$ is a nonsingular symmetric $N\times N$ matrix.\footnote{\, More generally, there can be a vector of regularization parameters, and $\Delta$ need not be nonsingular for $\Gamma$ to be nonsingular, but such additional intricacies do not change the conclusions drawn herein, so we will keep things simple.} Next, we discuss what the inverse of $\Gamma$ looks like.

\subsection{Deformed Covariance Matrix}

{}Let $V_i^{(a)}$ be $N$ right eigenvectors of $C_{ij}$ corresponding to its eigenvalues $\lambda^{(a)}$, $a=1,\dots,N$:
\begin{equation}
 C~V^{(a)} = \lambda^{(a)}~V^{(a)}
\end{equation}
with no summation over $a$. Let $U$ be the $N\times N$ matrix of eigenvectors $V^{(a)}$, {\em i.e.}, the $a$th column of $U$ is the vector $V^{(a)}$:
\begin{equation}
 U_{ij} \equiv V_i^{(j)}
\end{equation}
Let $\Lambda$ be the diagonal matrix of the eigenvalues $\lambda^{(a)}$:
\begin{equation}
 \Lambda_{ij} \equiv \delta_{ij} ~\lambda^{(j)}
\end{equation}
with no summation over $j$. Then
\begin{equation}
 C = U~\Lambda~U^T
\end{equation}
Note that, because $C$ is symmetric, $U$ can be chosen to be orthonormal: $U^T~U = 1$.

{}Let $J$ be the subset of large eigenvalues $\lambda^{(j)}$, $j\in J$. Let $J^\prime$ be the subset of small eigenvalues. (Note that $J \cup J^\prime = \{1,\dots,N\}$ and $|J| = M$.)\footnote{\, If some alphas were exactly 100\% (anti-)correlated, then $|J|$ would be smaller than $M$. For the sake of simplicity, here we are assuming that no alphas are exactly 100\% (anti-)correlated.} Let (no summation over $j$)
\begin{eqnarray}
 &&{\widetilde \Lambda}_{ij} \equiv \delta_{ij}{\widetilde \lambda}^{(j)}\\
 &&{\widetilde \lambda}^{(j)} \equiv \lambda^{(j)},~~~j\in J\\
 &&{\widetilde \lambda}^{(j)} \equiv 0,~~~j\in J^\prime
\end{eqnarray}
{\em I.e.}, ${\widetilde \Lambda}$ is obtained by setting the small eigenvalues in ${\Lambda}$ to zero. Let
\begin{equation}\label{C-tilde}
 {\widetilde C} \equiv U~{\widetilde \Lambda}~U^T
\end{equation}
Since the small eigenvalues are due to computational rounding, we can use ${\widetilde C}$ instead of $C$ in the definition (\ref{Gamma}) of $\Gamma$:
\begin{equation}\label{Gamma1}
 \Gamma\equiv {\widetilde C} + \epsilon~\Delta
\end{equation}
Note that ${\widetilde C}$ is exactly singular.

{}However, $\Delta$ is nonsingular and can also be decomposed as follows:
\begin{equation}
 \Delta = X~Z~X^T
\end{equation}
where $X^T~X = 1$ and
\begin{equation}
 Z_{ij} \equiv \delta_{ij}~v_j
\end{equation}
is a diagonal matrix of the eigenvalues $v_i$ of $\Delta$, which is assumed positive-definite.

\subsection{Inverse $\Gamma$}\label{sub3.2}

{}Note that
\begin{eqnarray}
 &&\Gamma = X~{\widetilde \Gamma}~X^T\\
 &&{\widetilde \Gamma}\equiv \epsilon~Z + {\widetilde \Omega}~{\widetilde \Omega}^T\\
 &&{\widetilde \Omega} \equiv X^T~\Omega
\end{eqnarray}
and $\Omega$ is an $N \times M$ matrix (recall that $|J| = M$) defined as follows (no summation over $A$):
\begin{eqnarray}
 &&\Omega_{iA} \equiv U_{iA}~\sqrt{{\widetilde \lambda}^{(A)}}\\
 &&A\in J
\end{eqnarray}
We have
\begin{equation}
 \Gamma^{-1} = X~{\widetilde \Gamma}^{-1}~X^T
\end{equation}
and
\begin{equation}
 {\widetilde \Gamma}^{-1} = \epsilon^{-1}~Z^{-1} - {\epsilon}^{-2}~Z^{-1}~{\widetilde \Omega}~Q^{-1}~{\widetilde \Omega}^T~Z^{-1}
\end{equation}
where $Q^{-1}_{AB}$ is the inverse of $Q_{AB}$ defined as
\begin{eqnarray}
 &&Q_{AB} \equiv \delta_{AB} + {\widetilde Q}_{AB}\\
 &&{\widetilde Q} \equiv \epsilon^{-1}~{\widetilde \Omega}^T~Z^{-1}~{\widetilde \Omega}
\end{eqnarray}
In the limit $\epsilon\rightarrow 0$ we have:
\begin{equation}
 \Gamma^{-1} = \epsilon^{-1}~\left[\Delta^{-1} - \Delta^{-1}~\Omega~(\Omega^T~\Delta^{-1}~\Omega)^{-1}~\Omega^T~\Delta^{-1}\right] +{\cal O}(1)
\end{equation}
In fact, the eigenvalues ${\widetilde \lambda}^{A}$ of $C$ do not even enter. Indeed, let us restrict $U$ such that it is an $N\times M$ matrix: $U \equiv (U_{iA})$. Then we have
\begin{equation}
 \Gamma^{-1} = \epsilon^{-1}~\Theta +{\cal O}(1)
\end{equation}
where
\begin{equation}
 \Theta \equiv \Delta^{-1} - \Delta^{-1}~U~(U^T~\Delta^{-1}~U)^{-1}~U^T~\Delta^{-1}
\end{equation}
{\em I.e.}, in the small $\epsilon$ (near-singular) limit, to the leading order the inverse of $\Gamma$ is determined solely by the inverse of the regulator matrix $\Delta$ and the eigenvectors $U_{iA}$ of $C$ corresponding to its large eigenvalues. Furthermore, in this limit the weights are given by
\begin{equation}
 w_i = {\widetilde \xi}~\sum_{j=1}^N \Theta_{ij}~\alpha_j
\end{equation}
where ${\widetilde \xi}$ is a normalization constant fixed by the weight normalization condition (\ref{w}). So, the weights for $S \rightarrow$ max in the singular limit are controlled by the choice of the regulator matrix $\Delta$.

\subsection{Diagonal $\Delta$: Weighted Regression}

{}When $\Delta$ is diagonal, $S \rightarrow$ max in the singular limit reduces to a simple weighted regression. Indeed, let
\begin{equation}
 \Delta_{ij} = \delta_{ij}~v_j
\end{equation}
We then have
\begin{equation}
 w_i = {{\widetilde \xi}\over v_i}~\left(\alpha_i - \sum_{j = 1}^N {\alpha_j \over v_j}~\sum_{A,B = 1}^K U_{iA}~U_{jB}~{\widehat Q}^{-1}_{AB} \right) \equiv {{\widetilde \xi}\over v_i}~\varepsilon_i
\end{equation}
where
${\widehat Q}^{-1}_{AB}$ is the inverse of
\begin{equation}
 {\widehat Q}_{AB}\equiv \sum_{i = 1}^N {1\over v_i}~U_{i A}~U_{i B}
\end{equation}
Note that
\begin{equation}
 \sum_{i=1}^N w_i~U_{iA} \equiv 0
\end{equation}
In fact, $\varepsilon_i$ are the residuals of a weighted regression (with weights $1/v_i$) of $\alpha_i$ over $U_{iA}$ (without an intercept). For non-diagonal $\Delta$ we have a generalized matrix-weighted regression.

\subsection{A Simple Regularization}

{}A simple regularization is given by:
\begin{eqnarray}
 &&\Gamma \equiv (1 - q)~D + q~C \equiv q \Gamma_1\\
 &&D_{ij} \equiv \delta_{ij}~C_{jj}
\end{eqnarray}
When $q\rightarrow 1$, we have $\epsilon\equiv (1 - q) / q\rightarrow 0$. The inversion of $\Gamma_1$ then produces the above weighted regression with $v_i = C_{ii}$, {\em i.e.}, the weights are inverse variances.

\section{Factor Model}\label{fac.mod}

{}Looking at (\ref{C-tilde}), one recognizes a multi-factor model -- well, of a very special form, that is, as it has vanishing specific risk. Its deformed version, (\ref{Gamma1}), with diagonal $\Delta$, however, has non-zero specific risk. The factor loadings matrix is simply the $N \times M$ matrix $U_{iA}$, and the factor covariance matrix is a diagonal $M\times M$ matrix ${\widetilde \Lambda}_{AB} = \delta_{AB}~{\widetilde\lambda}^{(B)}$, $A,B\in J$ (see Subsection \ref{sub3.2} for definitions). The $M$ risk factors comprising the columns of the factor loadings matrix $U_{iA}$ are nothing but the first $M$ principal components of the covariance matrix $C_{ij}$. This is essentially all in the spirit of the APT risk model.

{}Here we can ask if we can construct more general multi-factor risk models for alpha streams. There are two main reasons for doing so. The off-diagonal elements of $C_{ij}$ are not expected to be particularly stable out-of-sample. This instability is inherited by the principal components and the factor loadings matrix $U$. Furthermore, here we have at most $M$ risk factors, which typically is small because the number of observations is limited for alpha streams -- including due to their ephemeral nature. So, we wish to increase the number of risk factors and improve their out-of-sample stability. How can we achieve this?

{}The key observation here is that, whatever we use as the risk factors, we cannot use the alpha correlations or correlations of other quantities with alphas to construct the factor loading matrix or compute the factor covariance matrix -- if we do this, we will not get much beyond the $M$ factors based on the principal components because the number of observations for alphas is limited to $M+1$. So, we need to build risk factors that are not based on correlations with alphas and for which we can compute the factor covariance matrix based on a number of observations, call it $M_F$, such that $M_F\gg M$, or use risk factors for which the factor covariance matrix is readily available one way or another. Before we discuss some ways of approaching this problem, let us set up our notations first.

\subsection{Generalities}

{}Just as in the case of a stock multi-factor risk model, instead of $N$ alphas, one deals with $F\ll N$ risk factors and the covariance matrix $C_{ij}$ is replaced by $\Gamma_{ij}$ given by
\begin{eqnarray}\label{GammaFM}
 &&\Gamma \equiv \Xi + \Omega~\Phi~\Omega^T\\
 \label{Xi}
 && \Xi_{ij} \equiv \xi_i^2 ~\delta_{ij}
\end{eqnarray}
where $\xi_i$ is the specific risk for each $\alpha_i$; $\Omega_{iA}$ is an $N\times F$ factor loadings matrix; and $\Phi_{AB}$ is the factor covariance matrix, $A,B=1,\dots,F$. {\em I.e.}, the random processes $\Upsilon_i$ corresponding to $N$ alphas are modeled via $N$ random processes $z_i$ (corresponding to specific risk) together with $F$ random processes $f_A$ (corresponding to factor risk):
\begin{eqnarray}\label{fac.1}
 &&\Upsilon_i = z_i + \sum_{A=1}^F \Omega_{iA}~f_A\\
 \label{fac.2}
 &&\left<z_i, z_j\right> = \Xi_{ij}\\
 \label{fac.3}
 &&\left<z_i, f_A\right> = 0\\
 \label{fac.4}
 &&\left<f_A, f_B\right> = \Phi_{AB}\\
 \label{fac.5}
 &&\left<\Upsilon_i, \Upsilon_j\right> = \Gamma_{ij}
\end{eqnarray}
Instead of an $N \times N$ alpha covariance matrix $C_{ij}$ we now have an $F \times F$ factor covariance matrix $\Phi_{AB}$, which is expected to be more stable out-of-sample. Assuming all $\xi_i > 0$ and $\Phi_{AB}$ is positive-definite, then $\Gamma_{ij}$ is also positive definite.\footnote{\, Strictly speaking, positive-definiteness of $\Gamma_{ij}$ does not require, {\em e.g.}, positive-definiteness of $\Phi_{AB}$, but considering the practical nature of our discussion here, we will not try to be most general.}

\subsection{Risk Factors}\label{sub4.2}

{}We have already discussed the principal component approach above. The question we wish to address is what other risk factors we can build for alphas. The analogy with the stock multi-factor models is a good starting point.

{}One approach to constructing a factor model for alphas is to have $F_{\rm{\scriptstyle{style}}}$ style risk
factors and $F_{\rm{\scriptstyle{cluster}}}$ cluster risk factors. In the case of stocks, cluster risk factors are
usually referred to as industry risk factors. Since here we are dealing with alphas,
we will refer to such risk factors as cluster risk factors. In the case of alphas, the
following style factors {\em a priori} appear to be appropriate: 1) volatility, 2) turnover,\footnote{\, Turnover roughly can be thought of as being analogous to ADDV (average daily dollar volume, which can be viewed as a measure of liquidity) to market capitalization ratio in the case of stocks. Because turnover typically is highly correlated with alpha's holding horizon (the shorter the holding horizon, the higher the turnover), including the turnover style risk factor in an alpha factor model affects the weighting of different holding horizon alphas when using such factor model in regressions and/or optimization (and this itself depends on how the turnover risk factor is defined). We discuss the effect of different holding horizons in more detail in Section \ref{concl}.\label{foot.14}}
and 3) momentum. One may wish to add other style factors depending on how
alphas are constructed, {\em etc}. One other (perhaps more difficult to implement) style
factor one may wish to consider is capacity,\footnote{\, Capacity roughly can be thought of as being analogous to market capitalization (or size) in the case of stocks.} {\em i.e.}, how much capital each alpha can
absorb on its own; this requires modeling impact ({\em i.e.}, nonlinear trading costs). We comment on capacity in Appendix \ref{app.a}.

{}In the case of stocks, cluster factors are (usually) based on industry classification.
In the case of alphas, one can use a taxonomy of alphas, {\em i.e.}, one classifies alphas
according to how they are constructed -- if the required data is available, that is.
Out of thousands of alphas one may construct, many are very similar to each other
by construction. It is then clear that this similarity makes them more correlated,
just as stocks belonging to the same industry are more correlated. Just as in the
case of stocks, it therefore makes sense to treat clusters as risk factors and model
correlations between alphas based on such risk factors as opposed to computing them
directly based on a large number $N$ of the time series corresponding to individual
alphas. One difficulty with the alpha classification approach, however, is that the details of how each alpha is constructed must be known to those who build the factor model for alphas, and that is not always the case. Furthermore, in the case of stocks, the industry classification generally is a very stable construct -- companies do not tend to jump industries often.\footnote{\, Here we do not consider ticker de-listings, M\&As, ticker changes or new ticker additions as an ``instability". An instability in an industry classification would mean that it was based on some underlying aspects of companies that would make tickers change industries frequently. That would be a poorly constructed industry classification.} Alphas, however, are ephemeral by nature.

{}So, the style risk factors we described above are more-or-less easy to implement, but the alpha classification not so much. Unfortunately, the number of style risk factors is not large enough to make substantial difference compared with $M$ principal component risk factors, because $M$ can be substantial ({\em e.g.}, if the time series is based on daily alphas with a 1-year look-back). One straightforward way to increase the number of style factors is to break up each style factor into quantiles. If the number of quantiles $k$ is uniform over $F_{\rm{\scriptstyle{style}}}$ style risk factors, then this way we increase the number of such risk factors to $F_{\rm{\scriptstyle{style}}}^k$, which can be substantial. However, the {\em effective} number of the resulting style risk factors, while quite possibly larger than $F_{\rm{\scriptstyle{style}}}$, may not be as large as $F_{\rm{\scriptstyle{style}}}^k$ because of high correlations between various quantiles. Nonetheless, the quantile method is a simple way of squeezing more juice out of style risk factors.

{}Another thought is to use the well-established risk factors for stocks as risk factors for alphas, at least for those alphas whose underlying tradables are stocks.\footnote{\, Here one can use one's multi-factor risk model of choice, such as BARRA, Northfield, Axioma, Quantigic, SunGard APT, {\em etc.} One can also use industry classifications, {\em i.e.}, use industries (or equivalent groupings, sometimes referred to as sub-industries) as risk factors, {\em e.g.}, based on Bloomberg, GICS, ICB, {\em etc.}\label{RM}} If alphas themselves have no intrinsic risk management, then this is a sound approach -- in fact, one may very well wish to do risk management in this way. However, normally alphas are expected to be hedged against most risk factors, so in this case risk factors for stocks are already (essentially) ``factored" out of alphas.\footnote{\, In case they are not, one can get a relatively large number of such risk factors for which the factor covariance matrix is either readily available if one uses a multi-factor risk model for stocks, or easily computable based on multi-year look-backs for daily stock returns. We will comment on how to build the corresponding factor loadings matrices below. Also, some alphas intentionally may have risk exposure, and care is needed not to suppress such alphas inadvertently.}

\subsection{Underlying Tradables as Risk Factors}

{}However, even if risk factors for stocks are factored out, specific risks for stocks are not. So, the idea is to use the underlying tradables -- the stocks themselves -- as risk factors.\footnote{\, The underlying tradables need not be stocks. They can be any instruments. The idea applies all the same. Also, this idea has much broader applicability beyond alpha streams (see below).} One needs to quantify this, {\em i.e.}, we need to construct the factor loadings matrix and the factor covariance matrix. Here is one way of doing this.

{}For concreteness, let the underlying tradables be U.S. equities (this is not a critical assumption), so $A$ labels stocks in the universe traded by the combined alphas. In the 0th approximation the covariance matrix is diagonal: $\Phi_{AB} = v_A\delta_{AB}$, where $v_A$ are the historical {\em variances} for stock returns $R_A$. To compute $v_A$ even if, say, $F = 2,500$, there is no need to go back 10 years, because variances are substantially more stable than covariances. So, the look-back for computing $v_A$ can be much shorter, {\em e.g.}, monthly or annual. Beyond the 0th approximation $\Phi_{AB}$ is not diagonal. The off-diagonal elements themselves need to be modeled via a factor model approach for stocks; however, as we discussed above, these are readily available (see footnote \ref{RM}). No alpha correlations or correlations with alphas are needed to obtain $\Phi_{AB}$.

{}The next step is to identify $\Omega_{iA}$. As we mentioned above, the information about how individual alphas are constructed may not be available to us. However, the position data for each alpha had better be available to us if we are to trade them. Let this position data be $P_{iAs}$, which is the dollar holding of the alpha labeled by $i$ in the stock labeled by $A$ at time labeled by $t_s$, normalized so that $\sum_A |P_{iAs}| = 1$ for each given pair $i,s$. We need to construct $\Omega_{iA}$ from $P_{iAs}$ -- assuming there is no other data available to us, that is. {\em I.e.}, we need to get rid of the time series index $s$. The obvious choice\footnote{\, The sum over $s$ can be, {\em e.g.}, monthly or annual for alphas with overnight holdings and shorter for intraday alphas (see below).} $\Omega_{iA} \equiv \sum_s P_{iAs}$ does not work as the sign of $P_{iAs}$ flips over time frequently (assuming alphas have short holding periods). Basically, $\Omega_{iA} \equiv \sum_s P_{iAs}$ is essentially as unstable as $\mbox{Cor}(\alpha_i, R_A)$, which cannot produce more than $M$ independent risk factors. It is clear that we need an {\em unsigned} quantity to define $\Omega_{iA}$. We can use
\begin{equation}\label{modulus.1}
 \Omega_{iA}\equiv \sum_s \left|P_{iAs}\right|
\end{equation}
This is no longer unstable or similar to $\mbox{Cor}(\alpha_i, R_A)$. However, there is a regime where this definition may not work or may produce the effective number of risk factors $F_1$ lower\footnote{\, $F_1$ is the number of nonzero (or, in practice, ``non-small") eigenvalues of the matrix $\Omega~\Phi~ \Omega^T$.} (or even substantially lower) than $F$. If most alphas are trading most stocks at their trading bounds most of the time, and if these bounds are essentially uniform, then it is clear that in this case most $\Omega_{iA}$ defined this way will be close to each other -- the extreme case being $\Omega_{iA} \equiv \gamma$, where $\gamma$ is independent of $i$ and $A$, in which case we would have only a single risk factor proportional to a unit vector (a.k.a. intercept). The in-between case is where the effective number of risk factors based on (\ref{modulus.1}) is $1 < F_1 < F$. If $F_1 \ll F$, then most bounds are saturated, so we can still keep the $F_1$ risk factors based on (\ref{modulus.1})\footnote{\, In practice, when mixing non-uniformly defined risk factors, one must deal with the issue of how to define the factor covariance matrix for such mixed factors, including relative normalizations between non-uniformly defined risk factors. In this regard, sometimes it is simpler to have a uniformly defined set of risk factors.}  and add more risk factors based on a quadratic invariant. We cannot use any covariances w.r.t. $s$, so we can choose:
\begin{equation}\label{var}
 \Omega_{iA}\equiv \sqrt{\mbox{Var}\left(P_{iAs}\right)}
\end{equation}
where Var is the variance w.r.t. $s$ for each given pair $i,A$. There is an alternative definition
\begin{equation}\label{var.modulus}
 \Omega_{iA}\equiv \sqrt{\mbox{Var}\left(\left|P_{iAs}\right|\right)}
\end{equation}
which, however, is not expected to make a huge difference. Furthermore, one can use MAD instead of $\sqrt{\mbox{Var}}$, but these are minor details, which are not going to make it or break it. So, with (\ref{modulus.1}), (\ref{var}) and/or (\ref{var.modulus}), one should be able to capture the $F$ risk factors, or a substantial number of them much greater than $M$. And this can be done for alphas with overnight holdings and purely intraday alphas (as well as alphas that receive substantial contributions both from overnight holdings and intraday realized P\&L). When dealing with intraday (components of) alphas, the variances in (\ref{var}) and (\ref{var.modulus}) are understood as appropriately defined intraday.

{}We have been cavalier with the normalization of $\Omega_{iA}$ as defined in (\ref{modulus.1}), (\ref{var}) and/or (\ref{var.modulus}). This is because once $\Omega_{iA}$ are identified, they need to be properly normalized anyway in order to construct the specific risks, which will complete the risk model. This is a nontrivial step, which we will not delve into here.\footnote{\, In fact, this step is the key ingredient to building a successful factor model and is treated as proprietary knowhow by factor model providers, including Quantigic.}

{}The risk factors in (\ref{modulus.1}), (\ref{var}) and (\ref{var.modulus}) are defined this way because the premise is that, whatever the risk factors one defines, the factor covariance matrix $\Phi_{AB}$ is either readily available or computable without using the alpha correlations or correlations with alphas\footnote{\, Using alpha variances is fine as they are much more stable than covariances and, unlike covariances, do not limit the number of risk factors to $M$.}. Basically, assuming the only information available is the position data $P_{iAs}$, there is not much of a choice in defining $\Omega_{iA}$.

{}Also, let us emphasize that the idea of using underlying tradables as risk factors applies beyond alphas. If we have any $N$ processes $X_{is}$, $i=1,\dots,N$ determined by $F$ processes $Y_{As}$, $A=1,\dots,F$ via $X_{is} = \sum_{A=1}^F P_{iAs} Y_{As}$, where $P_{iAs}$ are previsible, then we can use $Y_{As}$ as risk factors for $X_{is}$ so long as the $P_{iAs}$ data is available to us. And this need not even be in the context of trading or finance.\footnote{\, {\em E.g.}, health risk factors, {\em etc.}}

{}Let us summarize by giving an outline for constructing risk factors for alphas. To simplify things, let us not mix different definitions of the factor loadings matrix. Let us focus on the definition (\ref{var}) uniformly across all alphas. This gives the factor loadings {\em up to an overall normalization factor}. The factor covariance matrix $\Phi_{AB}$ then is just the covariance matrix for stocks labeled by $A=1,\dots,F$. This covariance matrix $\Phi_{AB}$ can itself be modeled as a factor model (see footnote \ref{RM}). In a simpler approximation one could use a diagonal $\Phi_{AB} = v_A\delta_{AB}$, where $v_A$ are stock return variances. Also, if the risk management is done at the level of individual alphas, then one may wish to remove from the definition of the alpha risk factors the linear combinations of stocks corresponding to the stock risk factors. {\em I.e.}, in this case the factor loadings matrix $\Omega_{iA}$ is replaced by another factor loadings matrix $\Omega^\prime_{iA^\prime}$, where $A^\prime = 1,\dots,F^\prime$, and $F^\prime = F - F_S$, where $F_S$ is the number of stock risk factors. On the other hand, if the risk management is not done at the level of individual alphas, then, as mentioned above, one can use the stock risk factors themselves as the alpha risk factors. Let $\Lambda_{Aa}$ be the factor loadings matrix for the stock risk factors, where $a=1,\dots,F_S$. Then we can define $P_{ias}\equiv \sum_{A = 1}^F P_{iAs}~\Lambda_{Aa}$, and use this $P_{ias}$ as above to define the corresponding $F_S$ alpha risk factors $\Omega_{ia}$ (up to a normalization factor).

{}Finally, let us note that we can combine the aforementioned alpha factor loadings matrices $\Omega_{iA}$ or $\Omega_{ia}$ (or $\Omega^\prime_{iA^\prime}$ plus $\Omega_{ia}$) with alpha style risk factors as well as the principal component risk factors, {\em etc.}, if desired. One then needs to deal with the aforementioned issue of proper relative normalization of non-uniformly defined risk factors. Another issue is that such non-uniformly defined risk factors generally have nonzero correlations. Even if the factor covariance matrix for each set is known, the factor covariance matrix across the sets is not necessarily known. Here one can take a factor model approach and treat each set as a ``supercluster" and compute the factor covariance matrix between the ``superclusters" using the alpha covariance matrix. Thus, let us assume that we have two (properly normalized) sets of risk factors $\Omega^{(1)}_{iA_1}$ and $\Omega^{(2)}_{iA_2}$ with the (properly normalized) factor covariance matrices $\Phi^{(1)}_{A_1 B_1}$ and  $\Phi^{(2)}_{A_2 B_2}$, respectively, with $A_1,B_1 =1,\dots,F_1$, and $A_2, B_2 =1,\dots,F_2$. Let
\begin{eqnarray}
 &&f^{(1)} \equiv \sum_{i=1}^N\sum_{A_1=1}^{F_1}\alpha_i~\nu^{(1)}_{A_1}~\Omega^{(1)}_{iA_1}\\
 &&f^{(2)} \equiv \sum_{i=1}^N\sum_{A_2=1}^{F_2}\alpha_i~\nu^{(2)}_{A_2}~\Omega^{(2)}_{iA_2}
\end{eqnarray}
where $\nu^{(1)}$ and $\nu^{(2)}$ are some weights -- we can choose, {\em e.g.}, equal weighting. Then the covariance between $f^{(1)}$ and $f^{(2)}$ is given by
\begin{equation}
 \left\langle f^{(1)}, f^{(2)}\right\rangle = \sum_{i,j = 1}^N \sum_{A_1 = 1}^{F_1}\sum_{A_2 = 1}^{F_2} C_{ij}~\Omega^{(1)}_{iA_1}~\Omega^{(1)}_{iA_2}~
 \nu^{(1)}_{A_1}~\nu^{(2)}_{A_2}
\end{equation}
where $C_{ij}$ is the sample alpha covariance matrix computed using the actual alpha time series. Since here we have a single (or a few, in case we have a few more risk factor sets) covariance and $M \gg 1$, this covariance can be acceptably stable. Then, {\em up to an overall normalization factor}, instead of the 0th approximation $\Phi_{A_1 A_2} = 0$, in the first approximation we can set $\Phi_{A_1 A_2} \approx \left\langle f^{(1)}, f^{(2)}\right\rangle$. Note that this method can be used in the case of quantiled style factors we discussed above as long as the number of such quantiled style factors $F_{\rm{\scriptstyle{style}}}^k$ is not larger (and preferably -- for stability reasons -- is much smaller) than $M$. In this case instead of two sets as above we have $F_{\rm{\scriptstyle{style}}}^k$ sets.\footnote{\, And this method also applies to the style risk factors themselves, when $k=1$.} To reiterate, in the above discussion the (relative) normalizations of the factor loadings matrices corresponding to non-uniformly defined factor loadings can be tricky, and they are fixed when the specific risks are computed.\footnote{\, Computation of the specific risks, which, as mentioned above, is a proprietary topic, is outside of the scope of this paper.}

\section{Concluding Remarks}\label{concl}

{}Our primary motivation for considering factor models for alphas streams is twofold (see Section \ref{fac.mod}). First, the off-diagonal elements
of the sample covariance matrix $C_{ij}$ are not expected to be particularly stable out-of-sample. Second, when $M < N$ (recall that $M+1$ is the number of observations in the alpha time series based on which $C_{ij}$ is computed), which typically is the case in practice including due to the ephemeral nature of alphas, then $C_{ij}$ is singular with only $M$ nonzero eigenvalues. In fact, in most practical applications $M\ll N$.

{}In this regard, using alpha factor models -- at least, for the purposes of alpha weight allocation (via regression or optimization) -- is warranted when the number of alphas $N$ is large. Indeed, we must have $F\ll N$ (recall that $F$ is the number of risk factors). If one is dealing with, say, a dozen or so alphas, then for any reasonable historical track record (that one would feel comfortable with to trade these alphas moving forward), the sample covariance matrix is non-singular. Furthermore, in such a case replacing the sample covariance matrix with a factor model covariance matrix based on, say, half a dozen risk factors would appear to be an overkill. If, for some reason, it is imperative to have $F<N$ risk factors, in the case of such small $N$ one can simply take the first $F$ principal components of the sample covariance matrix as such factors. Anything else would appear to be overly contrived.\footnote{\, In this regard, there is a similarity with equities: if one is trading a couple of dozen stocks, equity factor models are not useful (at least, not in their applications that we focus on here). When one is trading, say, 1,000-2,500 stocks, then using factor models for equities is warranted.}

{}Furthermore, a large number of almost 100\% correlated alphas would imply that the true number of independent alphas is much lower. While it is relatively straightforward to construct a large number of highly correlated alphas ({\em e.g.}, one can take simple mean-reversion and/or momentum alphas and tweak parameters), it is much harder to construct a large number of alphas with low correlations. It takes a large number of quant researchers and developers and substantial hardware capabilities (for data mining purposes) to construct a large number of not-too-correlated alphas. All such efforts are highly proprietary, so it is not practicable to cite applications or tests of our framework and methods to/on real-life alphas.\footnote{\, And not because this has not been done.}

{}The purpose of this paper is to set up a general framework for constructing factor models for alpha streams with the view of elucidating it as a viable possibility. It appears that many practitioners might not have even considered factor models for alphas streams, likely due to the fact that usually practitioners view more familiar factor models for equities such as BARRA, Northfield, Axioma, {\em etc.}, as something one acquires from a provider, not something one builds in-house. However, both equities and alpha factor models can be built in-house. In fact, in the case of factor models for alphas there is no other choice but to build them in-house, as the required information relating to the alphas is highly proprietary and often rather specific from shop-to-shop in what details of alphas are available.\footnote{\, {\em E.g.}, as mentioned in Section \ref{sub4.2}, in some cases information may be available to create alpha taxonomy, in others only the position data may be available, which is insufficient for taxonomy.} As ``standardized" alpha factor models appear unlikely, custom alpha factor models are the way to go.

{}In footnote \ref{foot.14} we mentioned the effect of the turnover risk factor on alphas with different holding horizons. For literature on combining signals with different half-lives, see \cite{PO29,PO30,PO36}. In the case of alphas, the holding horizons are related to turnover: the shorter the holding
horizon, the higher the turnover. However, there is another practical consideration. Typically, low turnover strategies have lower Sharpe ratio and lower return on capital and higher capacity than higher turnover strategies. Higher Sharpe ratio/return strategies often enjoy higher performance fees, while their lower turnover counterparts can have substantially lower performance fees, but high capacity yields large dollar amounts in management fees. Because of these considerations, in practical applications mixing strategies with vastly different turnovers usually is not simply a question of weight allocation. On the other hand, weighting alphas with varying but in the same ballpark turnovers can be handled through appropriately defining the turnover style risk factor.\footnote{\, {\em E.g.}, one can define it via a log of turnover, further conformed to a normal distribution. Alternatively, one can, say, choose to suppress high turnover strategies (to mitigate trading costs) even more and define the turnover risk factor without taking a log.} However, one must also incorporate trading costs into optimization (or regression), which requires accounting for turnover reduction -- when alphas are combined, some trades are crossed, and the resultant portfolio turnover is reduced. Turnover reduction significantly differentiates alpha portfolio optimization problem from stock portfolio optimization problem, but the former is still tractable.\footnote{\, See \cite{AlphaOpt,AlphaWeights}, which utilize the spectral model of turnover reduction of \cite{SpMod}.} In this regard, holding horizons affect alpha weights, among other ways, via trading costs.

\appendix

\section{Capacity}\label{app.a}

{}In Subsection \ref{sub4.2} we mentioned capacity as a possible risk factor. Here we discuss capacity briefly. If there are no costs or only linear costs are present, portfolio capacity is unlimited. Once we introduce nonlinear costs (impact), portfolio capacity has a finite bound. Capacity is simply the value of the investment level $I = I_*$ at which the P\&L (computed for optimized alpha weights) is maximized. Let $\alpha\equiv\sum_{i=1}^N\alpha_i~w_i$ for the optimized weights. The P\&L is given by
\begin{equation}
 P = \alpha~I - L~D - {1\over n}~Q~D^n = T~\left({\widetilde M}~I - {1\over n}~Q~T^{n-1}~I^n\right)
\end{equation}
where $D = I~T$ is the dollar amount traded, $L$ is the linear cost per dollar traded, $T$ is the turnover, the impact coefficient $Q$ and power $n>1$ are model-dependent (and can be measured empirically), and
\begin{equation}
 {\widetilde M} \equiv {\alpha \over T} - L
\end{equation}
is the effective ``profit margin", which includes linear costs (but not impact).

{}The P\&L is maximized at
\begin{equation}
 I_* = {1\over T}~\left({{\widetilde M}\over Q}\right)^{1\over{n-1}}
\end{equation}
Note that the capacity $I_*$ increases as the turnover $T$ decreases. At the capacity bound we have the following P\&L
\begin{eqnarray}
&&P_* = M_*~T~I_* = {{n-1}\over n}~\left({{\widetilde M}^n\over Q}\right)^{1\over{n-1}}\\
&&M_*\equiv {{n-1}\over n}~{\widetilde M}
\end{eqnarray}
Here $M_*$ is the ``profit margin" at capacity. Note that for $n=1.5$, which is often assumed, we have $M_* = {\widetilde M}/3$, $I_* = {\widetilde M}^2/T Q^2$ and $P_* = {\widetilde M}^3/3Q^2$. Note that ${\widetilde M}$ depends on the turnover $T$.

{}When we combine a large number $N$ of alphas, we can use the spectral model of \cite{SpMod} to model turnover reduction due to the crossing of trades, according to which in the leading order in the $1/N$ expansion the portfolio turnover is given by
\begin{equation}
 T \approx \rho_*~\sum_{i=1}^N T_i ~\left|w_i\right| \equiv \rho_*~\tau
\end{equation}
where $T_i$ are individual alpha turnovers, and $0<\rho_*\leq 1$ is the turnover reduction coefficient, which can be computed using Eq. (34) of \cite{SpMod}.
With the same caveats as in the case of the turnover, one can repeat the arguments of \cite{SpMod} for the impact coefficient $Q$ and argue
that in the large $N$ limit we also have $Q \approx \rho_*~\kappa$, where $\kappa \equiv \sum_{i=1}^N Q_i \left|w_i\right|$, and $Q_i$ are the impact coefficients for individual alphas. Then the $\rho_*$ dependence of the capacity bound is given by
\begin{equation}\label{Istar}
 I_*\approx {1\over \tau~\rho_*^{n\over{n-1}}}~\left({\widetilde M}\over \kappa\right)^{1\over{n-1}}
\end{equation}
For $n=1.5$ we have $I_*\sim 1/\rho_*^3$. Recalling from \cite{SpMod} that $\rho_*$ is {\em roughly} an average correlation between the alphas, (\ref{Istar}) gives the power-law for the capacity dependence on the average correlation of alphas in a portfolio.

{}In general $n$ needs to be measured empirically. Measuring it directly for a portfolio is difficult. Below we give a simple method for measuring $n$ for {\em individual stocks}. We can then approximate $n$ for a portfolio as a weighted (these weights can be uniform) average of the impact powers $n_A$ for individual stocks.

\subsection{Measuring Impact}

{}The following discussion, unless otherwise stated, applies to an individual stock. The notations in this subsection are self-contained and should not be confused with the notations in the rest of the paper. $M$, with an appropriate index, denotes midquote. $P$, with an appropriate index,
denotes last print. Let us partition a trading day into $N$ equal intervals $I_i$ spanning
time between $T_{i-1}$ and $T_i$, $i = 1,\dots,N$, where $T_0$ = 9:30 AM, and $T_N$ = 4:00 PM.\footnote{\, One can choose these intervals to mimic real-life executions, {\em e.g.}, how VWAP is executed. In particular, they need not be uniform.}

{}For each interval $I_i$, let $P_{ia}$ ($a = 1,\dots,K_i$) be all prints at times $T_{ia}$, where
$T_{i-1} \leq T_{ia} < T_i$. Let $V_{ia}$ be the corresponding volumes traded. Let $A_{ia}$ and $B_{ia}$ be
the ask and bid prices at the times $T_{ia}$. For the sake of simplicity, we will exclude
all times $T_{ia}$ with $A_{ia} \leq B_{ia}$. That is, in the following, unless otherwise stated,
summation over $a = 1,\dots,K_i$ is understood to exclude datapoints with crossed and locked markets ($A_{ia} \leq B_{ia}$).

{}Next, let
\begin{equation}
 W_{ia} \equiv F\left(2~{{P_{ia} - B_{ia}}\over{A_{ia} - B_{ia}}}\right)
\end{equation}
where $F(x) \equiv \mbox{sign}(x)~\mbox{min}(|x|, 1)$. $W_{ia}$ act as weights.
For $P_{ia} = A_{ia}$ we have $W_{ia} = 1$, for $P_{ia} = B_{ia}$ we have $W_{ia} = -1$,
and the weight is 0 if the print is at the midquote. The following method can also
be implemented with simplified weights where one uses $\mbox{sign}(W_{ia})$ instead of $W_{ia}$.\footnote{In general, one can utilize variations of the method described here. One can also look at bid
and ask sizes and the (properly weighted) order book depth and add other bells and whistles, including $A_{ia} \leq B_{ia}$ cases, {\em etc}.}

{}Now we define
\begin{eqnarray}
 && V_i \equiv \sum_{a=1}^{K_i} W_{ia}~V_{ia}\\
 && U_i \equiv M_i - M_{i-1}
\end{eqnarray}
Here $U_i$ is the change in the stock price (or, more precisely, in its midquote) during
the interval $I_i$. We use midquotes at the endpoints of the interval as opposed to the
prints because the prints have extra noise in them, {\em e.g.}, due to the fact that prints
can occur at different pricepoints within the same bid-ask spread.

{}The impact model discussed earlier in this appendix assumes that the nonlinear cost of trading scales with the traded volume $V$ as
$|V|^n$. Here $V > 0$ for shares bought and $V < 0$ for shares sold. The impact on the
price then scales as $n~\mbox{sign}(V)~|V|^{n-1}$, which is the first derivative of $|V|^n$ w.r.t. $V$.
Our goal is to measure the power $n$.

{}This can be achieved by modeling impact using the datapoints $(V_i, U_i)$ defined
above. First, we exclude all such datapoints with $\mbox{sign}(U_i) \neq \mbox{sign}(V_i)$. Then in
R notations $n - 1$ can be determined as the coefficient of $\ln|V|$ in the linear model
(with an intercept)
\begin{equation}
 \ln|U_i|\sim\ln|V_i|
\end{equation}
The datapoints used in the linear model can span different trading days.

\subsubsection{Executions}

{}The above method is based on intraday pricing and volume data and allows to
estimate the expected value of $n$ on average, without taking into account the actual
executions in a given strategy. Better or worse executions could lead to a different
realized value of $n$, which we will denote as $\nu$. The challenge with determining $\nu$ is
that it is difficult to determine the price change analogous to $U_i$ above attributable
solely to executions in a given strategy because this strategy is only one of many
market participants affecting the price. Instead, what can be determined is the cost
$C$ of trading $D$ dollars. Again, as above, one can break up the trading day into
intervals $I_i$ and calculate the actual cost of trading $D_i$ dollars by comparing the fill
prices $F_{ia}$ ($a = 1,\dots, K_i$) for the corresponding numbers of shares $V_{ia}$ during the
interval $I_i$ with the midquote $M_{i-1}$ at the beginning of such interval. Then we have
\begin{eqnarray}
 &&C_i = \sum_{a=1}^{K_i} V_{ia}~\left(F_{ia} - M_{i-1}\right)\\
 &&D_i = M_{i-1}~\left|\sum_{a=1}^{K_i} V_{ia}\right|
\end{eqnarray}
Here one can use a benchmark other than $M_{i-1}$, as applicable. Also, summation over $a = 1,\dots,K_i$ is not restricted here.

{}The cost $C$ is modeled as
\begin{equation}
 C = L~D + {1\over\nu}~Q~D^\nu
\end{equation}
However, here we have three unknowns, linear slippage\footnote{\, For our purposes here, $L$ includes only linear slippage and excludes fixed trading costs (such as SEC fees, exchange fees, broker-dealer fees, {\em etc.}).} $L$, the impact coefficient $Q$ and
the power $\nu$. To circumvent this, one can scan values of $\nu$ (around the value of $n$,
if the latter has been measured as outlined above) and fit the coefficients $L$ and
$Q$ via a linear model. Thus, let ${\widetilde Q}\equiv Q/\nu$. Then in R notations $L$ and ${\widetilde Q}$
are the coefficients of the linear model (without an intercept)
\begin{equation}
 C \sim -1 + D + I(D^\nu)
\end{equation}
One can then pick the value of $\nu$ corresponding to the best fit.
Finally, let us mention that the values of $n$ and $\nu$ measured as outlined above
will vary from stock to stock. For portfolios one can use a median or weighted
average value.

\section{Factor Covariance Matrix and Specific Risk}\label{app.b}

{}In this appendix we discuss a simple method for obtaining the factor covariance matrix for a set of risk factors based on the sample covariance matrix computed based on alpha time series. This approach is useful when the factor covariance matrix for a set of risk factors is not readily available or computable. An example of this kind of a situation is the case of style risk factors (or their quantiled versions discussed in Subsection \ref{sub4.2}. Another example is the case of alpha clusters based on alpha taxonomy -- if one can be built, that is.

{}In the factor model approach we have (\ref{fac.1})-(\ref{fac.5}) and (\ref{Xi}). Let
\begin{eqnarray}
 &&Q_{AB} \equiv \sum_{i=1}^N \Omega_{iA}~\Omega_{iB}\\
 &&{\widetilde Q}_{AB} \equiv Q^{-1}_{AB}
\end{eqnarray}
We have
\begin{equation}\label{PhiAB}
 \Phi_{AB} = \sum_{C,D=1}^F {\widetilde Q}_{AC}~{\widetilde Q}_{BD}~\left(\sum_{i,j=1}^N \Omega_{iC}~\Omega_{jD}~\Gamma_{ij} - \sum_{i=1}^N\xi_i^2~\Omega_{iC}~\Omega_{iD}\right)
\end{equation}
On the other hand, we have
\begin{equation}\label{xi.sq}
 \xi_i^2 = \Gamma_{ii} - \sum_{A,B=1}^F\Phi_{AB}~\Omega_{iA}~\Omega_{iB}
\end{equation}
Plugging (\ref{xi.sq}) into (\ref{PhiAB}) we get a matrix equations for $\Phi_{AB}$:
\begin{equation}\label{Phi-T}
 \Phi_{AB} - \sum_{C,D,C^\prime,D^\prime=1}^F {\widetilde Q}_{AC}~{\widetilde Q}_{BD}~T_{C D C^\prime D^\prime}~\Phi_{C^\prime D^\prime} =
 \sum_{C,D=1}^F {\widetilde Q}_{AC}~{\widetilde Q}_{BD}~\sum_{i,j=1,~i\neq j}^N \Omega_{iC}~\Omega_{jD}~\Gamma_{ij}
\end{equation}
where
\begin{equation}
 T_{ABCD}\equiv\sum_{i=1}^N \Omega_{iA}~\Omega_{iB}~\Omega_{iC}~\Omega_{iD}
\end{equation}
is a totally symmetric 4-tensor. So, the idea is that, in cases where $\Phi_{AB}$ is not independently computable, one can fix it via (\ref{Phi-T}) by replacing $\Gamma_{ij}$ via the sample covariance matrix $C_{ij}$.

\subsection{Binary Factor Loadings}

{}The above discussion simplifies substantially if the factor loadings $\Omega_{iA}$ are binary, {\em i.e.}, they take only two values, 0 or 1, and indicate if the alpha labeled by $i$ belongs to the alpha cluster labeled by $F$:
\begin{eqnarray}
 &&\Omega_{iA} = \delta_{G(i), A}\\
 &&G:\{1,\dots, N\} \mapsto \{1,\dots, F\}
\end{eqnarray}
where $G$ is the map between alphas and the alpha clusters. Also,
\begin{equation}
 N_A \equiv \sum_{i=1}^N \delta_{G(i), A}
\end{equation}
is the number of alphas that belong to the cluster labeled by $A$. Note that
\begin{equation}
 \sum_{A=1}^F N_A = N
\end{equation}
We have:
\begin{eqnarray}
 &&Q_{AB} = N_A~\delta_{AB}\\
 &&{\widetilde Q}_{AB} = {1\over N_A}~\delta_{AB}~\\
 &&T_{AAAA} = N_A~~~(\mbox{other components vanish})\\
 &&\Phi_{AB} = {1\over N_A~N_B}~\sum_{i:G(i) = A}~\sum_{j:G(j) = B} C_{ij},~~~A\neq B\\
 &&\Phi_{AA} = {1\over N_A~(N_A-1)}~\sum_{i,j:G(i) =A,~G(j) = A,~i\neq j} C_{ij}\\
 &&\xi_i^2 = C_{ii} - \Phi_{G(i),G(i)}
\end{eqnarray}
Here one can immediately see the issue we mentioned in Section \ref{fac.mod}: $\xi_i^2$ are not guaranteed to be positive.\footnote{\, As we mentioned in Section \ref{fac.mod}, the resolution of this issue is a proprietary topic, which is outside of the scope of this paper.}

\end{document}